# Observation of Alfvén solitons in the solar corona using Parker Solar Probe (PSP) and Solar and Heliospheric Observatory (SOHO)

Murchana Khusroo[1,2], Ankita Thapa[2]

Abstract

Solitons are predominantly observed in near-Earth plasmas as well as planetary magnetospheres; however, their existence in the solar corona remains largely unexplored, despite theoretical investigations. This study aims to address this gap by examining the presence and dynamics of solitons in the solar corona, particularly in the context of coronal heating. Utilizing observational data from the Parker Solar Probe (PSP) and the Solar and Heliospheric Observatory (SOHO) during the onset of a strong Coronal Mass Ejection (CME) event, the analyses reveal a train of aperiodic solitons with increasing amplitude preceding the eruption. A key finding of this study is that the observed aperiodic soliton train serves as a potential candidate for facilitating energy transfer through dissipation within the coronal plasma, thereby, influencing the initiation of solar eruptive events such as CMEs. A defining characteristic of this solitary train is its *hypersonic* $(M > 3)$ and *super-Alfvénic* $(M_A > 1.5)$ nature, as evident from the presence of high Mach numbers, which reinforce its role in plasma energy equilibration in the solar corona, thereby contributing to plasma heating.

[1] Corresponding author dr.mkhusroo@gmail.com
[2] Department of Physics, University of Science and Technology Meghalaya, Ri-Bhoi, Meghalaya, 793101, India

# 1. Introduction

Solitary structures, predominantly referred to as solitons, are ubiquitous in space plasmas and are frequently observed in the near-Earth plasma environments as well as in planetary magnetospheres, including the Earth's magnetosphere (Moola et al., 2003; Matsumoto, et al., 1994). Since Scott Russell's pioneering demonstration in 1844, solitons have become one of the most extensively studied phenomena in fluid dynamics, including plasma physics. Although solitons can be generated under controlled laboratory conditions and observed in various natural setting, space plasmas provide a unique natural environment where their detection has been significantly facilitated by advancements in space exploration technologies (Khusroo & Bora, 2019). In space plasmas, a soliton refers to a localized, self-reinforcing nonlinear wave packet that maintains its shape and size while propagating through the plasma, essentially acting like a solitary wave that is stable against disruptions. It is a localized wave pulse that arises from a delicate balance between nonlinearity and dispersion (Chen, 1984). Depending upon the nature of the plasma wave inducing the perturbation, solitons can manifest as either electrostatic or electromagnetic structures. In space plasmas, a soliton represents a fundamental class of nonlinear wave phenomenon, capable of propagating over a long distance while maintaining its shape and size. Electrostatic solitary waves (ESWs) are single wave pulses that are generated due to the accumulation of charge (electrons or ions) density in a particular region in space, triggered by nonlinear perturbation of the medium by an electrostatic plasma wave, such as an ion-acoustic wave or an electron plasma wave. When an inhomogeneity of charge distribution develops due to the evolution of nonlinear electrostatic perturbation in space, it leads to the formation of a potential structure at that particular region, thereby generating an electric field which is henceforth, detected by a spacecraft in the form of a bipolar pulse (Khusroo & Bora , 2015). On the contrary, an electromagnetic solitary wave emerges when the plasma is perturbed by an incident electromagnetic wave, such as an Alfvén wave, which manifests as magnetic field fluctuations detected by the spacecraft. These electromagnetic solitons are often observed in the Earth's magnetosphere in the form of an aperiodic train of solitons with either increasing or decreasing amplitudes (Khusroo & Bora, 2019); typically occurring during strong magnetic field fluctuations associated with intense events like geomagnetic storms (Sarma & Khusroo, 2024). These structures are recurrently observed in astrophysical plasmas, including solar wind plasmas (Yang et al.,

2024; Shah et al., 2020); however, empirical evidence of their existence in solar corona remains elusive.

ESWs have been widely investigated through theoretical frameworks and observational studies. Temporal and spatial localized electrostatic structures in the auroral regions were identified through Freja and FAST satellite missions in the 1990s, providing explicit empirical confirmation of their existence (Marklund et al., 1995). More recently, Cluster spacecraft observations at $\sim 5\,R_E$ have revealed localized structures with peak-to-peak electric field fluctuations of $\sim 30-170\,mV\,m^{-1}$ and lifetimes of $10-280\,s$ (Marklund et al., 2004; Khusroo & Bora, 2015). A notable fact about these nonlinear structures observed in the near-Earth plasmas is that they exhibit characteristic timescales corresponding to both ion-acoustic as well as Alfvénic modes, as substantiated by experimental observation and theoretical analyses (Ekeberg et al., 2010). A theoretical investigation by M Khusroo and M P Bora (2015) demonstrated that a negative electron thermal anisotropy ($T_{e\perp}/T_{e\parallel} > 1$) plays a crucial role in the formation of these large amplitude ESWs, with results closely aligning with observational data (Khusroo & Bora, 2015). More recently, a theoretical investigation on the role of solitons in a protoplanetary disk (PPD) shed light on how hypersonic electrostatic solitons, formed out of magnetorotational instability (MRI) - driven electron heating turbulence in the dead zones of a PPD, can serve as an efficient energy equilibration mechanism through soliton decay and radiation (Das, Khusroo, & Bora, 2025). Other studies on electromagnetic solitons have identified oscillations triggered by mirror instability in the planetary magnetospheres (Cattaneo et al., 1998) and cometary environments (Tsurutani et al., 1999), while K. Stasiewicz (2004), in his study, demonstrated that these trains of magnetic pulses are slow-mode magnetosonic solitons, emerging as intrinsic solutions of a Hall-magnetohydrodynamic (MHD) plasma model (Stasiewicz, 2004). In another theoretical study by M Khusroo and M P Bora (2019) on magnetospheric plasma dynamics within a Hall-MHD framework incorporating finite Larmor radius (FLR) effects, periodic oscillations with constant amplitude were ruled out. Instead, they predicted a train of magnetosonic solitons with evolving amplitudes, consistent with spacecraft observations of aperiodic oscillations in the magnetopause region (Khusroo & Bora, 2019). S Sarma and M Khusroo (2024) carried out an in-depth observational investigation of such aperiodic magnetospheric oscillations during the recovery phase of a geomagnetic storm in solar cycle 23, utilizing data from Cluster II mission that revealed a similar train of aperiodic solitons with progressively increasing amplitudes propagating through the bow shock region, highlighting the nonlinear evolution of solitary structures in response to magnetospheric perturbations induced by CMEs and ICMEs (Sarma & Khusroo, 2024). The presence of solitons in the solar wind plasma has also been investigated through observational data and numerical simulation, elucidating their contribution to the solar wind turbulence at 1 AU ( Shah et al., 2020).

In a seminal study, K. Stasiewicz (2006) demonstrated that the two fluid equations governing a collisionless plasma admit nonlinear solutions in the form of a dissipative Alfvén solitons termed as alfvenons. He proposed that alfvenons form as magnetohydrodynamic (MHD) perturbations that propagate upward through the solar corona, where the Alfvén speed decreases. Given their role in electromagnetic energy dissipation and plasma heating within planetary magnetospheres, these structures have been considered as a compelling candidate for explaining coronal heating (Stasiewicz, 2006; Stasiewicz & Ekeberg, 2008). However, the direct observation of solitons in the solar corona remains largely unexplored, with no conclusive evidence reported to date. Furthermore, most investigations on solitons have primarily focused on space and magnetospheric plasmas, while their existence and dynamics within the solar corona are yet to be examined from a solar physics perspective.

The coronal heating problem remains one of the fundamental unsolved mysteries in solar physics, referring to the unexplained rise in temperature from the Sun's photosphere (~ 6,000 K) to the corona, which reaches millions of Kelvin (Alfvén, 1941; Peter & Dwivedi , 2014). For decades, researchers have explored various mechanisms, including wave heating, magnetic reconnection, and turbulence-driven dissipation, to explain this extreme temperature disparity (Sturrock, 1999 ; Ayaz et al., 2024 ; Verdini et al., 2010; Kuperus et al., 1981). Observations from space missions like the SOHO, the Solar Dynamics Observatory (SDO) and the PSP, alongside theoretical and numerical modeling, have provided insights into potential energy transfer processes, yet a definitive explanation remains elusive (Sigalotti & Cruz, 2023; Telloni et al., 2023; Li et al., 2015; Huber & Lühe, 1991). As mentioned by Stasiewicz, the long-standing problem of coronal heating can be investigated through soliton perspective, as these nonlinear potential structures have the ability to transfer energy through radiation as well as dissipation (Stasiewicz, 2006). This study aims to bridge this gap by leveraging multi-spacecraft mission data to investigate soliton dynamics in the solar corona, offering new insights into their existence and behavior in this regime. In Section 2, we present the data and methodology employed in this investigation, focusing on a selected CME event and analyzing observational data from the Parker Solar Probe (PSP) and the Solar and Heliospheric Observatory (SOHO) missions. In Section 3, the results are highlighted with detailed illustrations. Section 4 provides a summary of the findings, while Section 5 concludes with a discussion of the implications.

## 2. Data and Methodology

The data used in this study to investigate and identify the Alfvén solitary structures or Alfvén solitons in the solar corona are retrieved from the FIELDS (Bale et al., 2016) instrument suite onboard PSP on 17 March 2023 from 18:09:00 UT to 18:13:00 UT just before the eruption of a Coronal Mass Ejection (CME), which was detected at around 19:29:06 UT-23:55:05 UT by the Large Angle and Spectrometric Coronagraph (LASCO) C2 (Brueckner et al., 1995) onboard SOHO, with the corresponding plots shown in Fig 1, plotted in Helioprojective coordinates (Thompson , 2006). The thermal plasma parameters are measured by the PSP Solar Wind Electron, Alphas, and Protons (SWEAP) instrument, including the Solar Probe ANalyzer (SPAN)-Ion payload ( Kasper et al., 2016), and the PSP position is measured in heliocentric coordinates using Helio1-day position data from SPDF HelioWeb for the same date (Wilkinson, 2012).

The magnetic field full-cadence data from the FIELDS instrument are first plotted for the extended time range of 18:00:00 UT to 23:00:00 UT to investigate variations in the solar corona (see Fig 2). A test case is then selected for the interval 18:09:00 UT to 18:13:00 UT, corresponding to the period immediately preceding the CME eruption. Minimum Variance Analysis (MVA) is performed on the magnetic field data during this interval, revealing a clear train of solitons with increasing amplitudes (Sonnerup & Cahill Jr., 1967; Dunlop, Woodward, & Farrugia , 1995). This analysis proves to be a powerful and effective method to identify these structures, complementing the traditional approach of wavelet analysis. It is to be noted that the FIELDS data are provided in RTN coordinates, where R (radial), T (Tangential), and N

(Normal) components correspond to the spacecraft-centric coordinate system (Dmitry, 2019). Throughout the manuscript and analysis, RTN coordinates are represented as XYZ components respectively, for simplicity and consistency in the data presentation. Wavelet transform (Meyer, 1992) and power spectral density (PSD) (Stoica & Moses, 2006) analysis are also applied to the magnetic field data that confirm the nonlinear oscillatory nature of the magnetic field and validate the presence of a train of Alfvén solitons with aperiodic amplitudes. The findings of these analyses are presented in Section 3.

3. Results

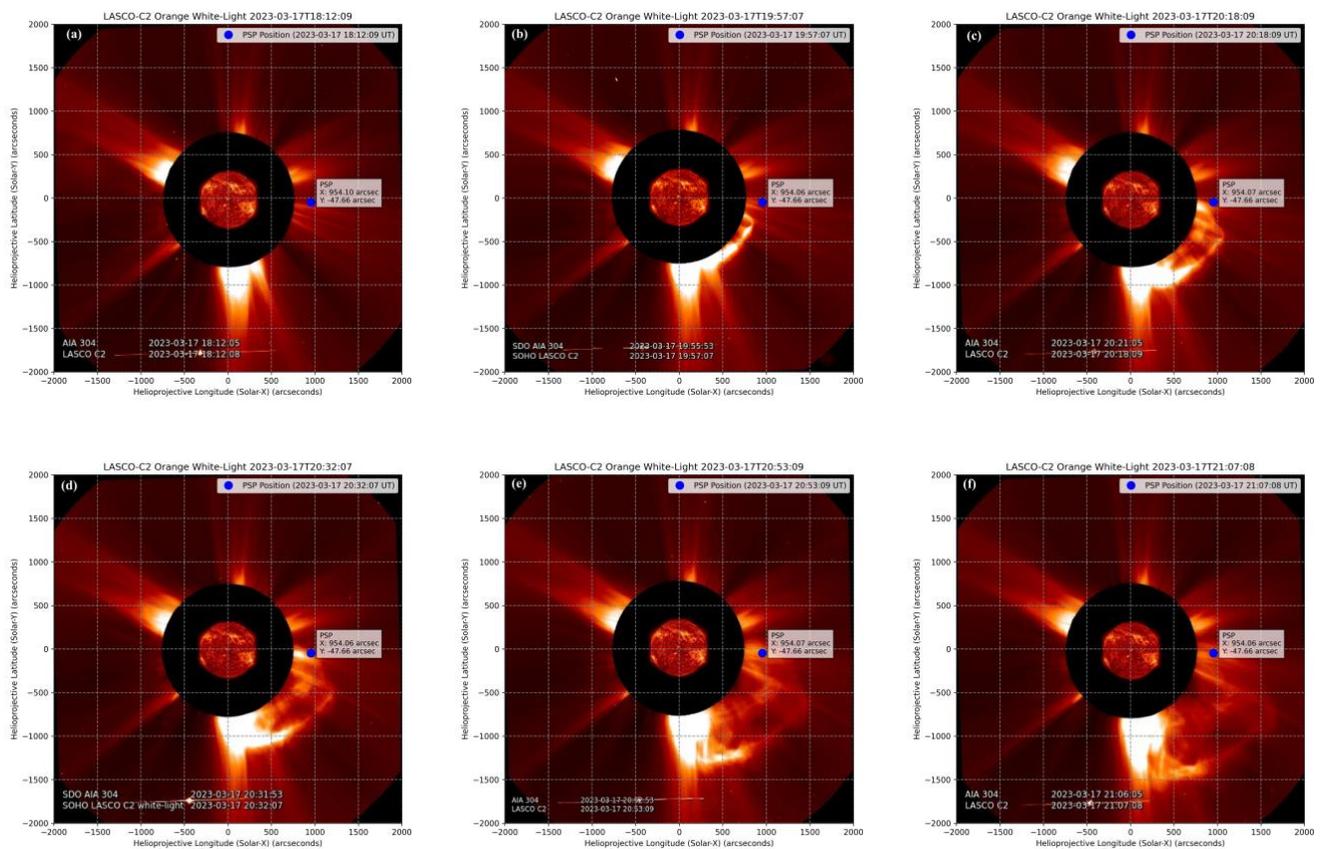

Fig 1: Evolution of a Coronal Mass Ejection (CME) detected by LASCO C2 Coronagraph and the position of PSP in Helioprojective coordinates from 18:00:00 UT to 23:00:00 UT on 17 March 2023.

The plot in Fig 2 on the left is a polar representation of the PSP's position in Stonyhurst Heliographic Coordinates on 17-03-2023 at 18:12:09 UT, generated using the PSP Helio1-day position data. These data specify the

PSP's location within the Stonyhurst Heliographic Coordinate system, a solar-centered reference frame with the Sun at its origin. On the right of the figure is a 2D scatter plot of the PSP's position from 16-03-2023 to 18-03-2023 in Helioprojective coordinates, expressed in arcseconds, with the date and time encoded in a color scale (Thompson , 2006). These plots confirm that the PSP was located well within the solar corona, which typically extends from 10 – 20 solar radii (approximately 0.0465 AU to 0.093 AU) (Johnson-Groh, 2021), during the formation of the solitary structures observed just before

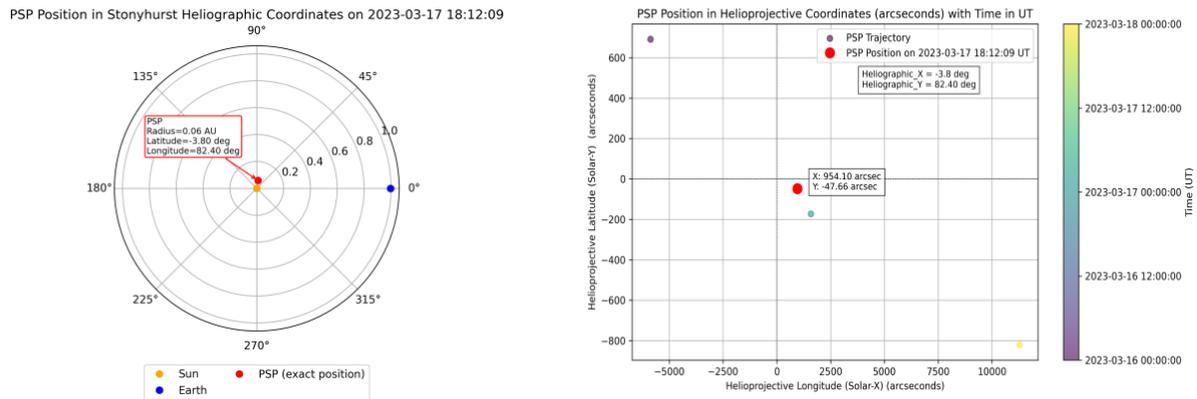

Fig 2: On the left is a polar plot showing the Parker Solar Probe's (PSP) position in Stonyhurst Heliographic Coordinates on 2023-03-17 at 18:12:09 UT. The plot is generated using Helio1-day position data from a Common Data Format (CDF) file, with the Sun at the origin. The PSP's position is marked with its heliographic latitude (-3.80°), longitude (82.40°), and radial distance (0.06 AU) from the Sun. The Earth is also shown for reference. On the right is a scatter plot of the Parker Solar Probe's position in helioprojective coordinates (arcseconds), highlighting the PSP's exact position on 2023-03-17 at 18:12:09 UT. The helioprojective coordinates are derived from the PSP's heliographic latitude and longitude, providing a detailed spatial view of its location relative to the Sun's center. The time evolution of the trajectory is indicated by a color gradient bar (in UT).

the eruption of the CME.

Fig 3 presents the magnetic field fluctuations (panel a), plasma and magnetic pressures (panels b & c respectively) and plasma beta, $\beta$ (panel d) from PSP that have been recorded from 18:00:00 UT to 23:00:00 UT, capturing the passage of CME-driven shock. The CME-shock is evident around 21:00:00 UT, marked by abrupt changes in all measured parameters. Panel (a) illustrates the magnetic field magnitude ($|B|$) that sharply decreases corresponding to the shock front. Panel (b) reveals distinct transitions in the RTN components of the magnetic field ($B_X$, $B_Y$, $B_Z$) in $nT$, highlighting significant variations dominated by $B_X$, indicative of the transition of plasma due to shock formation. Panel (c) displays the plasma pressure in $nPa$, with noticeable peaks up to $\sim 150\,nPa$, while panel (d) illustrates the variation in magnetic pressure showing relatively stable trends punctuated by a sudden dip coinciding with magnetic field $|B|$ fluctuations. Panel (e) plots the plasma beta ($\beta$), which is generally low, indicating magnetic pressure dominance, except for a sharp spike around 21:00:00 UT, where plasma pressure temporarily exceeds magnetic pressure due to the compression and heating associated with the CME shock.

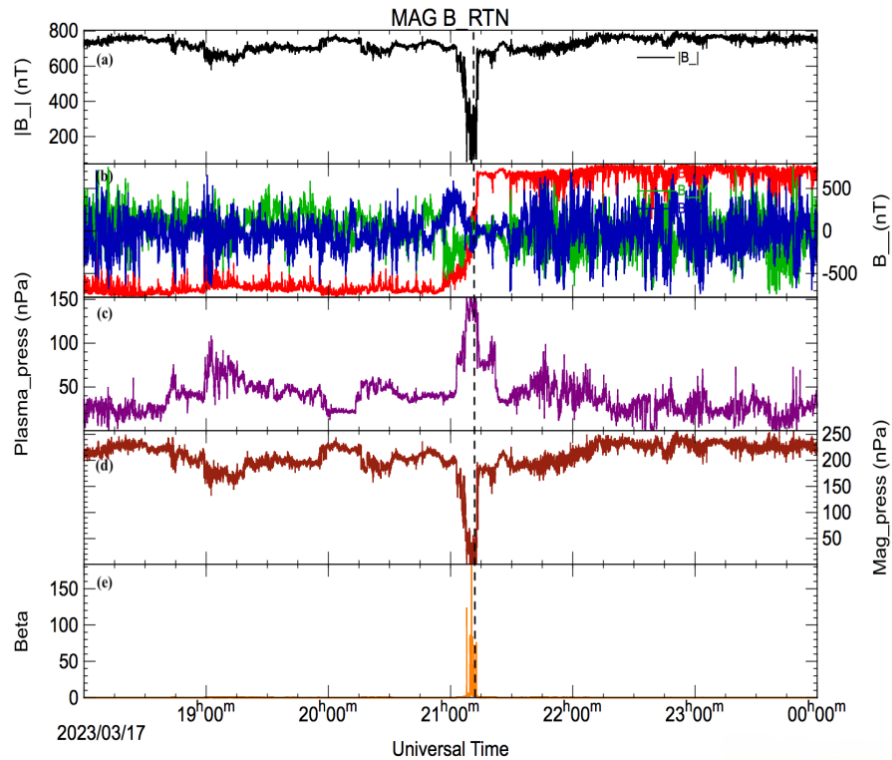

Fig 3: Temporal evolution of the magnetic field and plasma parameters observed by PSP for the CME event recorded by the LASCO C2 coronagraph between 18:00:00 UT and 23:00:00 UT on 17/03/2023. Panel (a) shows the total magnetic field magnitude ($|\boldsymbol{B}|$), panel (b) displays the magnetic field components in the RTN coordinate system with the red, green, and blue curves representing $\boldsymbol{B_X}$, $\boldsymbol{B_Y}$, and $\boldsymbol{B_Z}$ respectively. Panel (c) shows the plasma pressure, panel (d) shows the magnetic pressure and panel (e) displays the plasma beta ($\boldsymbol{\beta}$). The CME-driven shock is distinctly visible within this interval and is marked by the black dashed line, highlighting a sudden variation in these parameters.

Fig 4, on the other hand, captures the temporal evolution of the magnetic field and plasma parameters for the test case interval 18:09:00 UT to 18:13:00 UT. The observed plasma conditions, with low beta and highly fluctuating magnetic fields, depict a nonlinear plasma environment, which paves way for the formation of nonlinear plasma structures, particularly Alfvén solitons in the solar corona. In Fig 5, the plots are generated after performing the MVA over the magnetic field data for the selected test case interval. The hodograms illustrate the projection of the magnetic field components over different variance axes as well as their directional variations along the respective eigenvectors of the variance matrix. When analyzed separately, the MVA result distinctly reveals the variation of the magnetic field components in different variance directions as shown in Fig 6. Notably, the $B_z$ component exhibits the maximum variance, indicating the emergence of a train of solitary structures with progressively increasing amplitude. These structures are identified as aperiodic Alfvén solitons characterized by their non-uniform amplitudes, in the context of our study, which is later, elaborated with the wavelet analysis and power spectrum density analysis shown in Fig 7 and Fig 8 respectively.

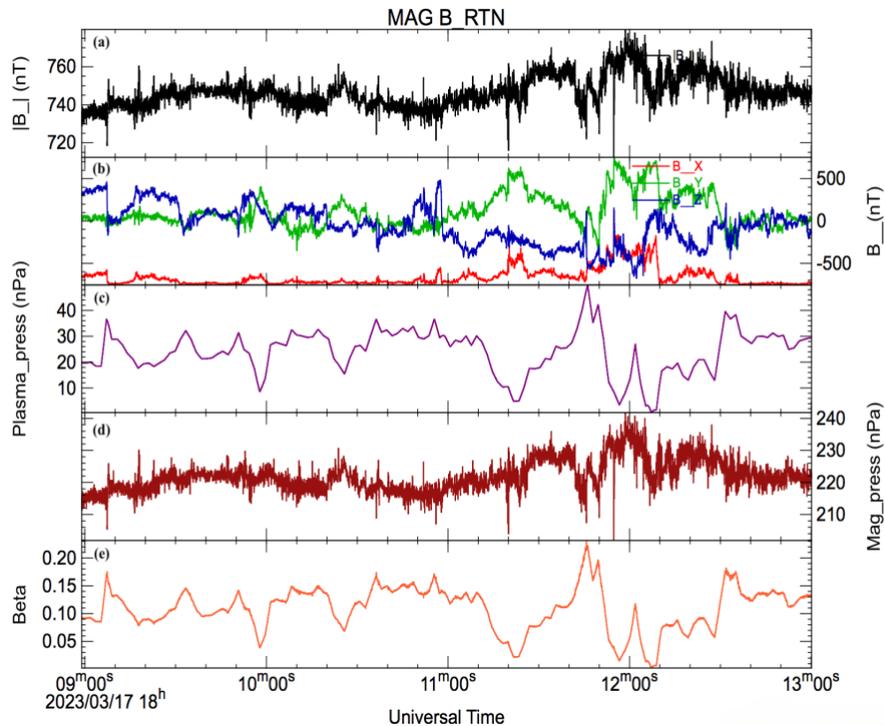

Fig 4: Compressed view of Fig 3, highlighting the test case interval from 18:09:00 UT to 18:13:00 UT.

Both figures illustrate the time-frequency distribution of magnetic field fluctuations performed through wavelet transform (WT) over the entire CME range (18:00:00 UT to 23:59:59 UT) as well as pre-CME range or the test case interval (18:09:00 UT to 18:13:00 UT), using two different wavelet filters, Haar (filter length 2) and Daubechies-4 (D4) respectively (Daubechies, 1992; Haar, 1910). The color scale indicates power intensity, with red and orange representing higher power. In Fig 7 the enhanced power is observed across multiple scales, particularly around the transition region at ~ 21:00:00 UT, highlighting the presence of potential structures such as solitons, alongside broadband turbulence and wave activity preceding the CME shock, marked by the blue dashed line. This is typically due to the nonlinear plasma processes arising from the stochastic behavior of the magnetic field lines in the solar corona during a solar eruptive event like a CME. Out of these nonlinear plasma phenomena, a train of aperiodic soliton formation is one such plasma phenomenon, which can contribute greatly to the generation of such immense power and energy because of its multifaceted nature, which is commonly observed and investigated in the other planetary environments (Williams et al., 2006; Dasgupta & Maitra, 2020). However, while the WT using the Haar filter effectively captures sharp discontinuities, it struggles to smoothly resolve gradual variations.

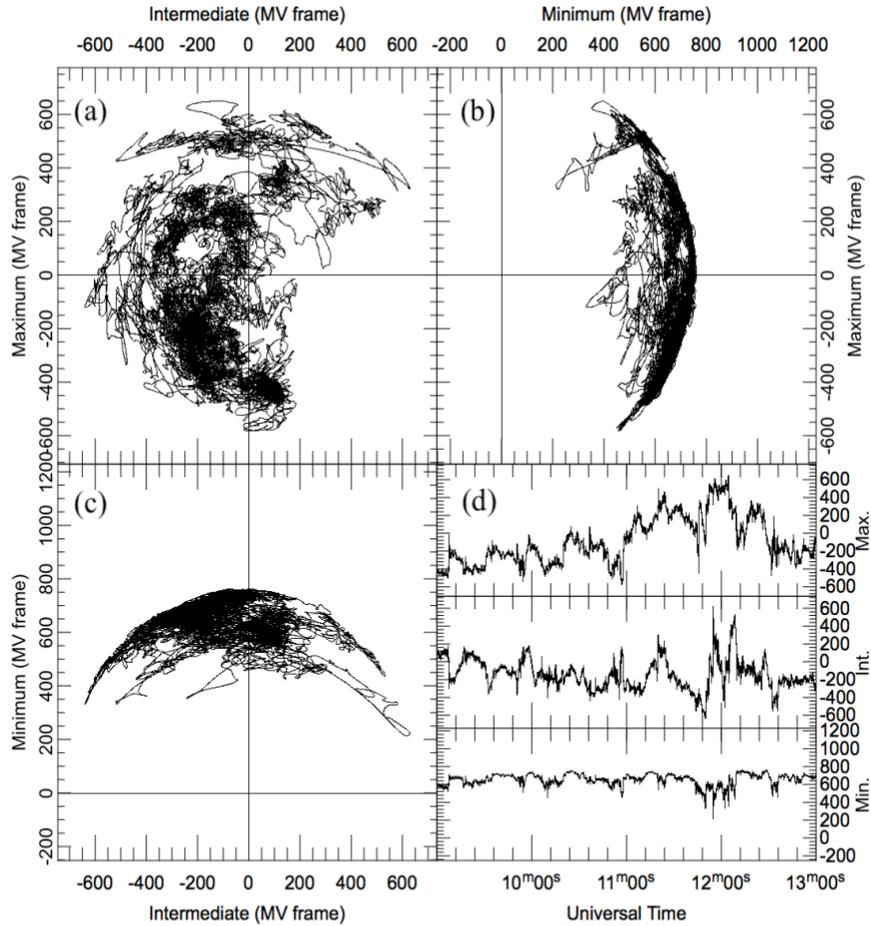

Fig 5: Variance analysis of the magnetic field components in the RTN frame for the time interval 18:09:00 UT to 18:13:00 UT. Panel (a), (b) and (c) shows hodograms of the magnetic field in the maximum, intermediate and minimum variance planes, with the axes representing the magnetic field fluctuations in the MVA frame. Panel (d) displays the time series of the magnetic field projections along the maximum, intermediate and minimum variance directions. The eigenvalues ($\lambda$) corresponding to the maximum, intermediate, and minimum variance directions are $7.01 \times 10^4$, $2.81 \times 10^4$ and $4.1 \times 10^3$, respectively with the associated eigenvectors listed in the bottom table.

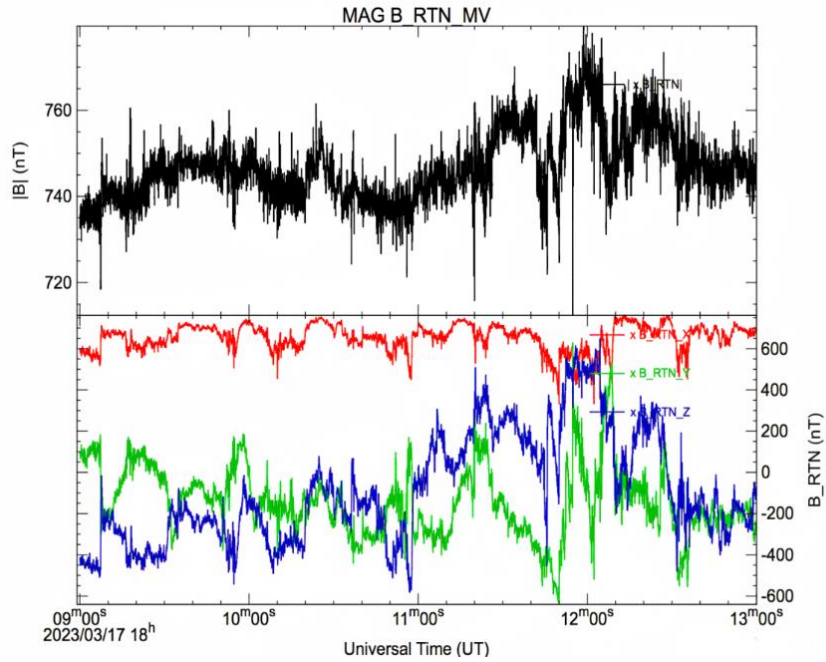

Fig 6: Magnetic field variations obtained after performing the MVA on the PSP data for the test case interval 18:09:00 UT to 18:13:00 UT. The top panel represents the magnitude of the magnetic field ($|B|$) in nanoTeslas (nT), showing significant fluctuations over time. The bottom panel illustrates the individual magnetic field components $B_X$ (red), $B_Y$ (green), and $B_Z$ (blue) along the principal variance axes.

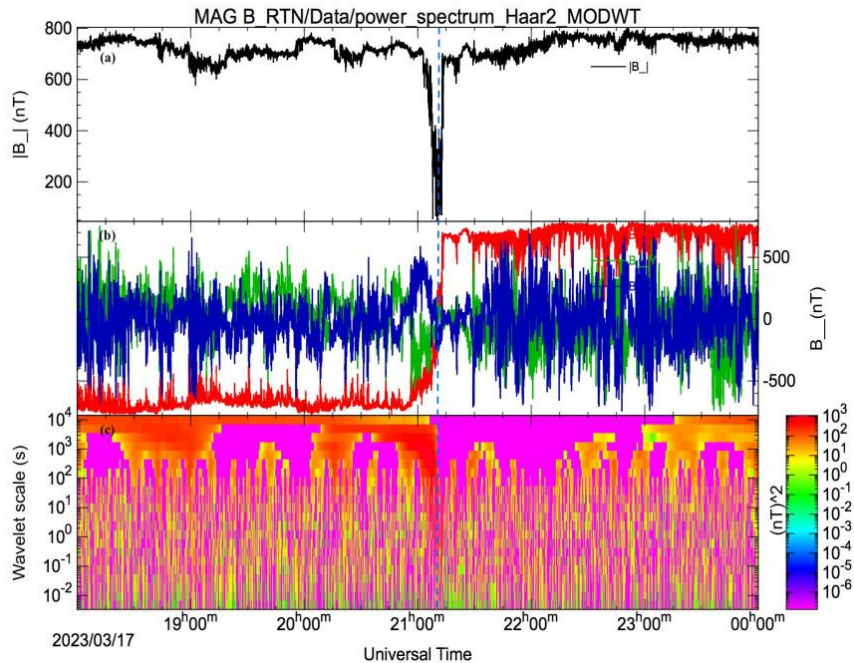

Fig 7: Wavelet transform of the magnetic field data from the PSP during and before the CME event recorded on 17 March 2023, using the wavelet filter Haar 2, with the Maximum Overlap Discrete Wavelet Transform (MODWT). The figure represents the Haar wavelet (order 2) multi-resolution analysis for the entire range of the CME from 18:00:00 UT to 23:59:59 UT, showing the magnetic field magnitude $|B|$ (panel a), the three components of the magnetic field in RTN coordinates (panel b), and the corresponding wavelet power spectrum (panel c), with color intensity indicating power variations at different time scales. The blue dashed line in the figure marks the CME shock, highlighting the sudden increase in power generation across multiple scales just before the CME eruption.

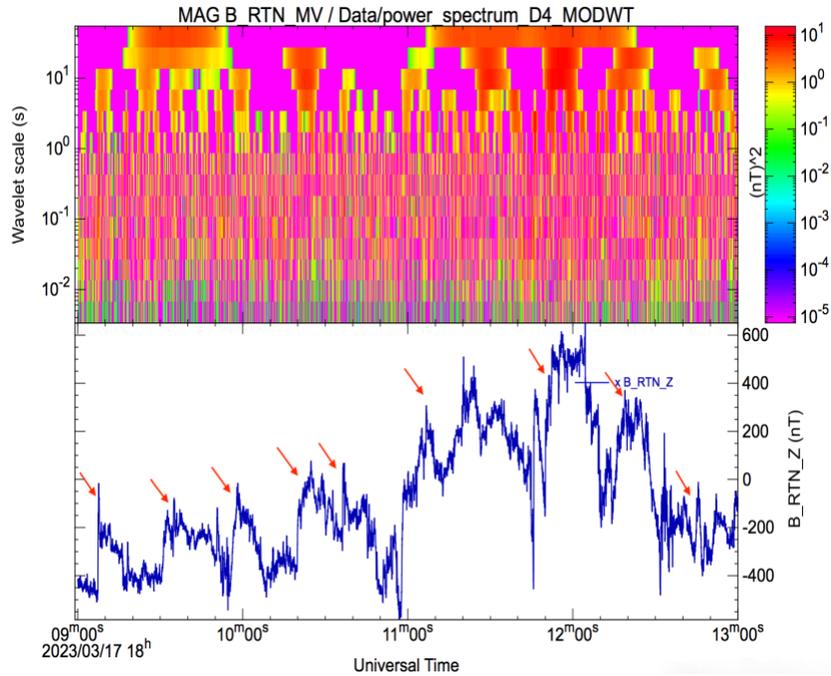

Fig 8: WT analysis using the Daubechies-4 (D4) wavelet filter for the test case interval from 18:09:00 UT to 18:13:00 UT, focusing on the $B_Z$ component of the magnetic field after performing the MVA as shown in Fig 6. The red arrows highlight soliton-like structures with increasing amplitudes observed in the time series.

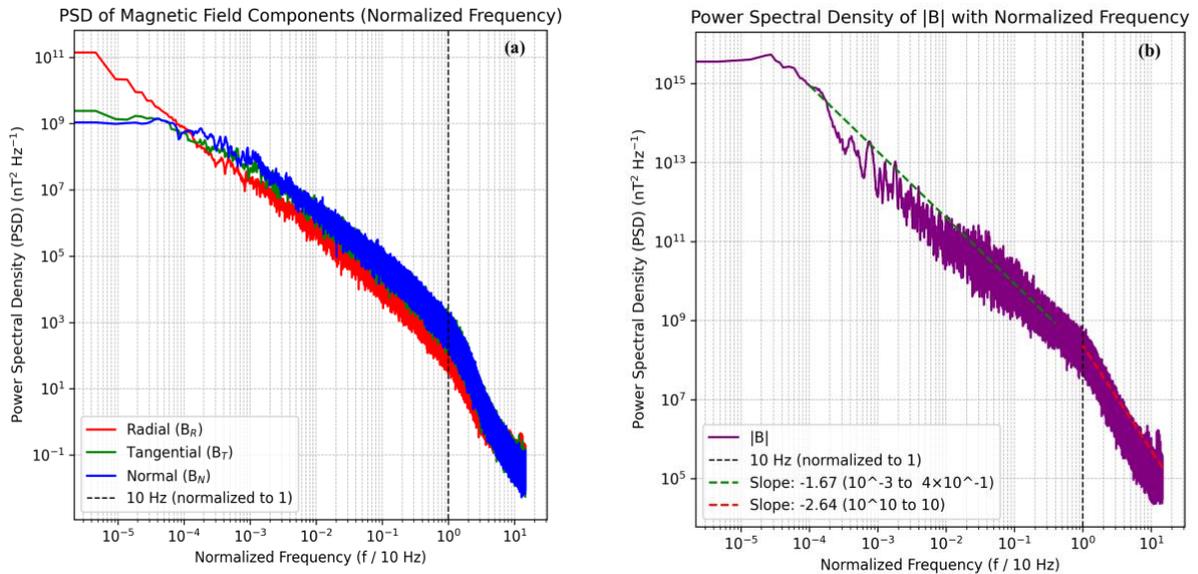

Fig 9: Power spectrum density (PSD) analysis of the magnetic field data from the PSP for the CME event recorded on 17 March 2023, from 18:00:00 UT to 23:00:00 UT as a function of normalized frequency. The frequency is normalized to 10 Hz in both cases. Panel (a) represents the PSD analysis of the magnetic field components in RTN coordinates with respective colors. Panel (b) shows the PSD analysis of the magnetic field magnitude $|B|$, with two distinct power-law slopes highlighted in green and red.

It mostly emphasizes abrupt variation in the magnetic field, such as shocks or discontinuities in plasma, where a strong, high-power feature is followed by a reduced, low- power region. The Daubechies D4 wavelet filter, on the other hand, is more efficient in providing smoother decompositions by resolving continuous oscillations. Fig 8 shows the wavelet decomposition of the data to analyze the solitonic structures. The presence of localized power regions in the wavelet spectrum clearly indicates the formation of Alfvén solitons in the solar corona just before the CME eruption.

To further analyze the soliton formation in the solar corona, a power spectral density (PSD) analysis has been performed over the entire range of the CME (18:00:00 UT to 23:59:59 UT) shown in Fig 9 (a and b). It is evident from the figure that the absence of any significant peak signifies the nonlinearity of the oscillations corresponding to the absence of any periodicity for any frequency '$f$'. This is another strong indication that these oscillations are a *group* of train of Alfvén solitons generated nonlinearly in the solar corona. In Fig 9 (b) the two distinct power-law slopes indicate different spectral regimes.

- The -1.67 green slope in the intermediate frequency range suggest the presence of turbulence with energy cascading at a rate different from the standard Kolmogorov turbulence (Kolmogorov, 1941).
- The −2.64 red slope at higher frequencies indicates a steeper spectral fall-off, potentially due to a dissipation process.

The dashed black line at 10 Hz (normalized to 1) marks a transition frequency, which corresponds to a characteristic wave mode or break in the turbulent cascade.

Fig 10 presents a detailed analysis of plasma parameters measured over the entire CME interval and Fig 11 shows the same plots for the test case interval. The plasma velocity components in the RTN coordinates for both cases are shown in Fig 12. The upper panel illustrates the variation in plasma velocity during the test case duration, ranging from 400 km/s to 600 km/s, with an estimated structure velocity of ~500 km/s.

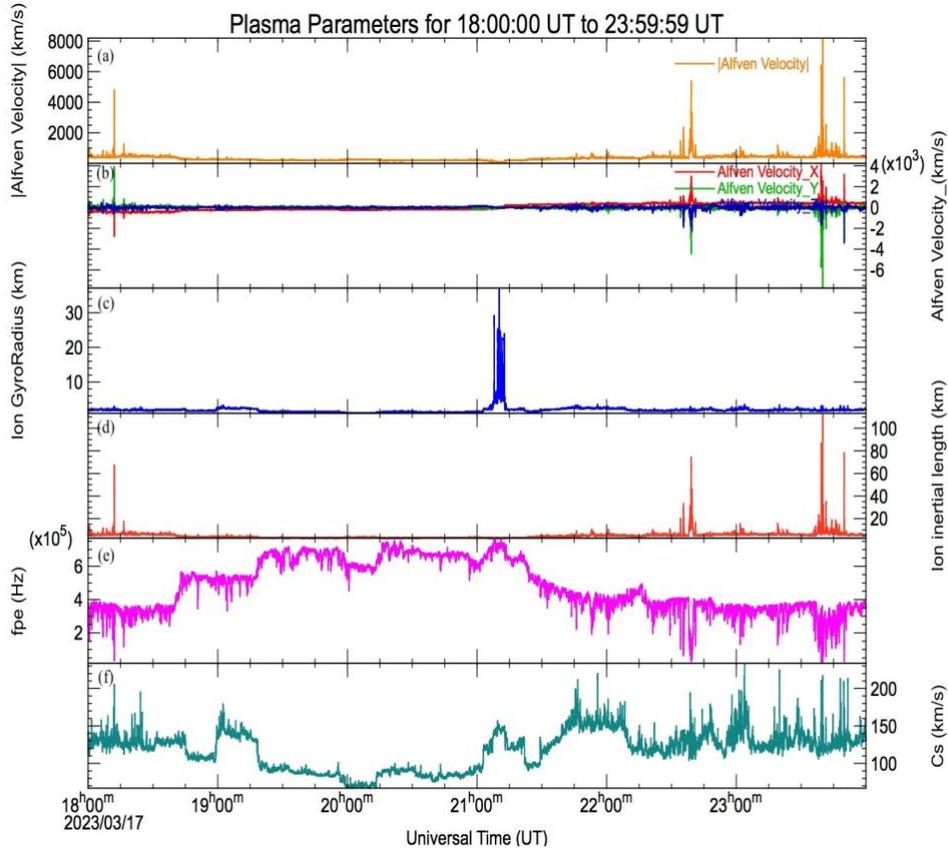

Fig 10: Temporal variations of the key plasma parameters observed for the CME event on 17 March 2023, from 18:00:00 UT to 23:59:59 UT. Panel (a) represents the magnitude of the Alfven velocity ($|V_A|$) in orange, while panel (b) presents the components of the Alfven velocity in RTN coordinates with respective colors. Panel (c) depicts the ion gyroradius while panel (d) shows the ion inertial length. Panel (e) displays the electron plasma frequency (fpe) in Hz and panel (f) represents the ion sound speed (Cs) in km/s.

Fig 13 (a) displays the density ($n_{p,i}$) and temperature ($T_{p,i}$) profiles of protons and He²⁺ ions respectively, in the coronal plasma, as measured by the SWEAP instrument onboard PSP for the entire CME range (18:00:00 UT to 23:59:59 UT). The increase in the density accompanied by a decrease in the temperature is quite notable. During this period, the density (blue curve) rises significantly, reaching values close to $7 \times 10^9 \, m^{-3}$, while the temperature (red curve) simultaneously decreases below $5 \times 10^5 \, K$. This inverse correlation is a typical nature commonly observed in CME-driven plasma, where the expanding magnetic field traps the dense plasma while reducing the thermal energy, leading to a temperature drop. Fig 13 (b) shows the variation of the Mach number (upper panel) and the Alfvén Mach number (lower panel) over the same duration. From the figure it is seen that the Mach number fluctuates predominantly between 2 and 4, with occasional spikes exceeding 5, indicating variations in plasma flow speed relative to the sound speed. Notably, the Alfvén Mach number (blue curve) remains relatively low throughout the duration but exhibits a sharp and transient spike around 21:00 UT, reaching values exceeding 20. This sudden increase in Mach number depicts the presence of the CME-shock, which is visible in Fig 3.

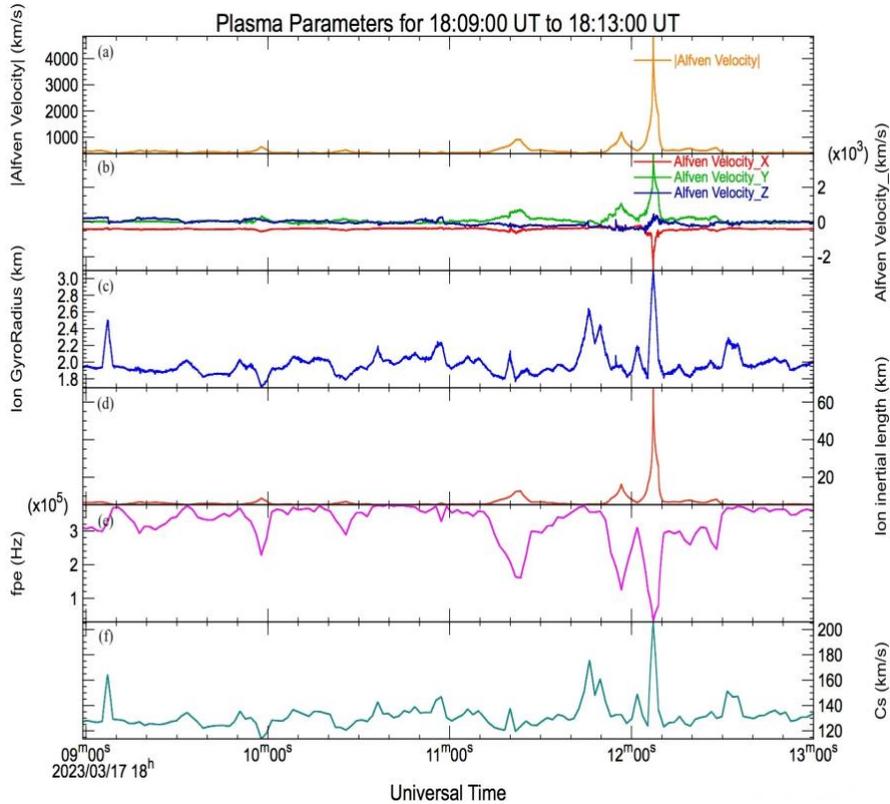

Fig 11: Plasma parameters for the test case interval in the pre-CME region from 18:09:13 UT to 18:13:00 UT.

For the test case interval, the similar variation of density, $n_{i,p}$ and temperature, $T_{i,p}$ along with Mach numbers is shown in Fig 14 (panels a and b respectively). The plot in panel (a) shows the density and temperature profile for the respective period while panel (b) displays the variation in Mach number and Alfvén Mach number for the same. From the figure it is seen that the Mach number (red curve) mostly remains above 3 ($M > 3$), with a sudden increase beyond 5 ($M > 5$) around 18:12:00 UT, indicating a strong enhancement in plasma velocity relative to the sound speed. Simultaneously, the Alfvén Mach number (blue curve) fluctuate around unity ($M \approx 1$), with a brief drop followed by a sharp rise beyond 1.2 ($M_A > 1.2$), suggesting variations in the

Alfvén speed likely caused by local plasma instabilities or wave interactions. The combination of these trends provides strong evidence supporting the presence of a soliton train interacting with the background solar wind just before the eruption of the CME.

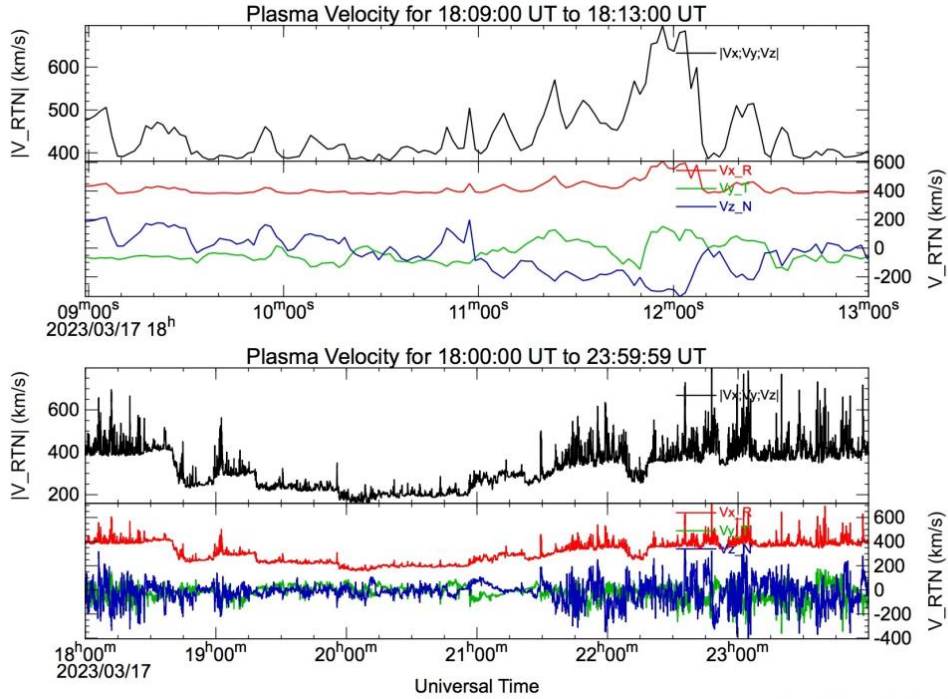

Fig 12: Variation of plasma velocity for both the cases. The top panel shows the variation for the test case while the bottom panel shows for the entire CME period.

This behavior aligns with the presence of solitons, which are capable of locally modifying the plasma properties. As solitons evolve, they can lead to soliton radiation, a process where energy is gradually transferred to smaller-scale fluctuations, ultimately contributing to turbulence or dissipation mechanisms in plasma (Murusidze et al., 1998; Das, Khusroo, & Bora, 2025). From the above analysis, it is clear that this aperiodic soliton train is a fast-moving nonlinear plasma structure or a fast magnetosonic aperiodic soliton train that is *hypersonic* and *super-Alfvénic* in nature. The plasma parameters as detected by PSP for both cases are given in Table 1.

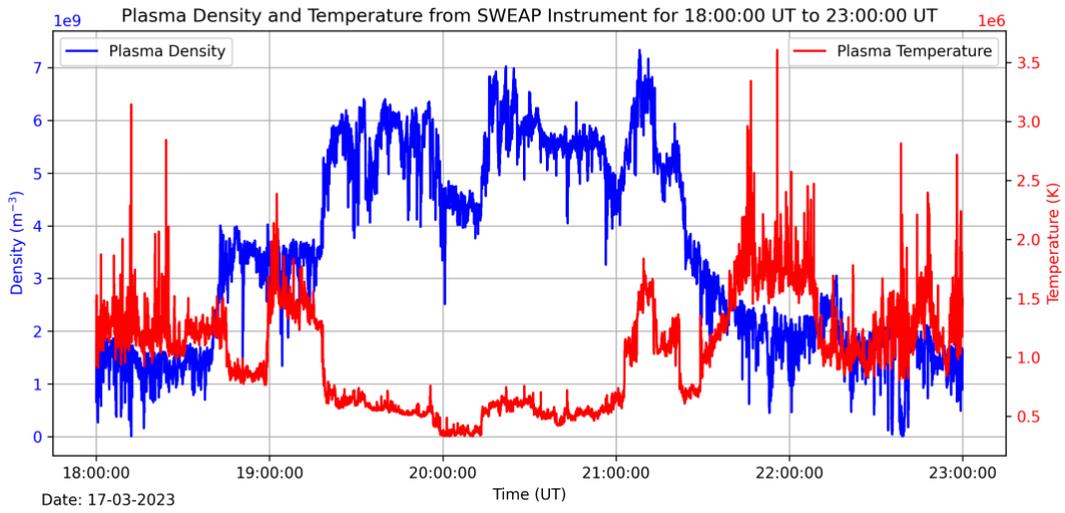

(a)

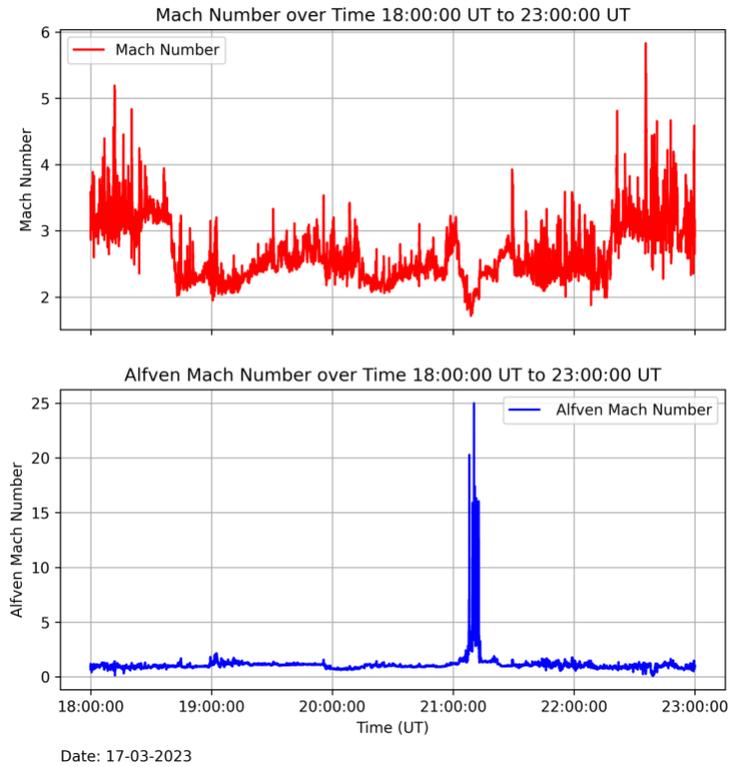

(b)

Fig 13: Plot (a) presents the temporal variation of density, $n_{p,i}$ and temperature, $T_{p,i}$ whereas plot (b) illustrates the temporal variation of Mach number and Alfven Mach number throughout the entire CME period.

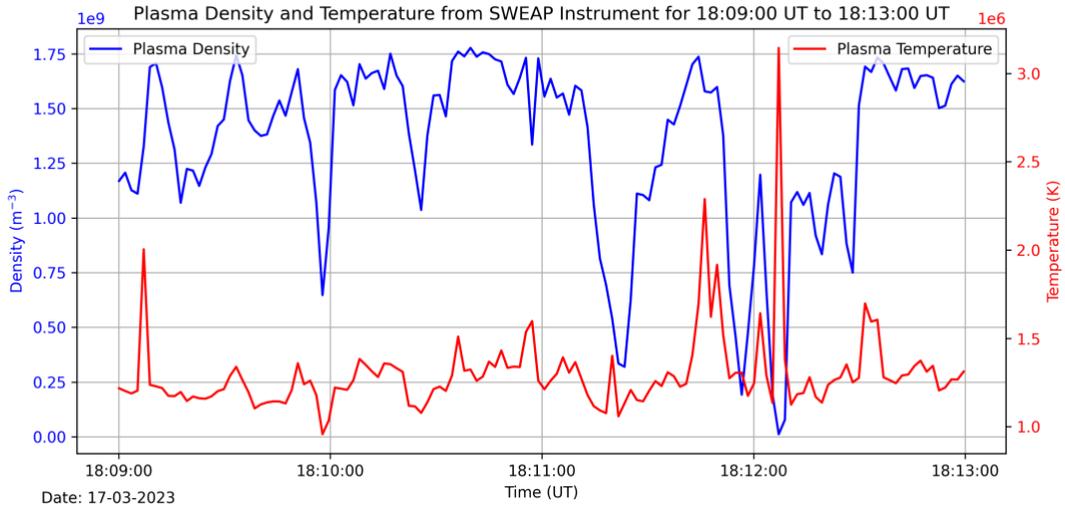

(a)

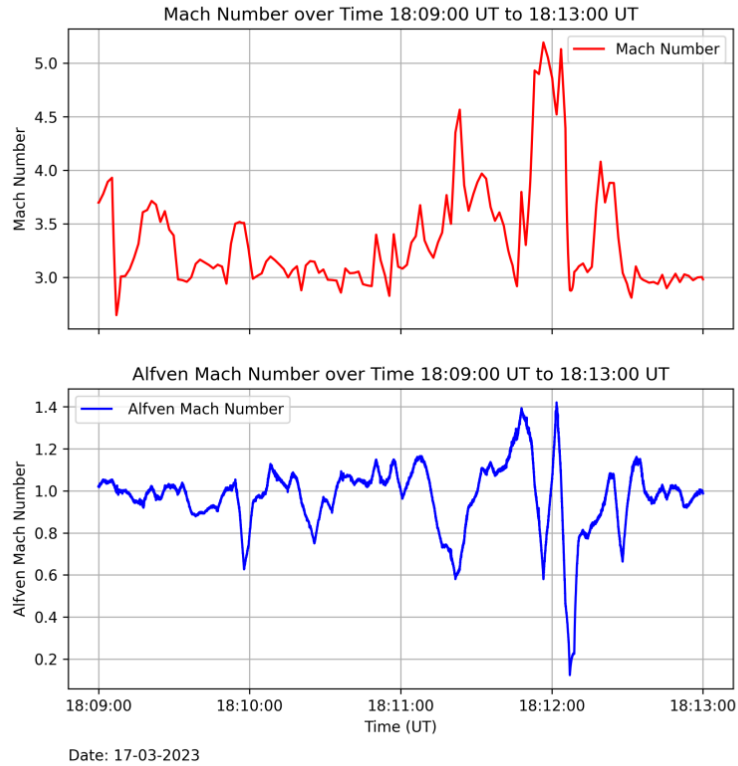

(b)

Fig 14: Plot (a) displays the temporal variation of density, $n_{p,i}$ and temperature, $T_{p,i}$ whereas plot (b) illustrates the temporal variation of Mach number and Alfven Mach number for the test case interval.

Table 1: Approximate values of plasma parameters for both the cases.

| Plasma Parameters | CME Event (18:00:00-23:59:59) UT | Soliton Train (18:09:00-18:13:00) UT |
|---|---|---|
| Magnetic Field, $B\ (nT)$ | $\approx 750$ | $\approx 750$ |
| Plasma Pressure $(nPa)$ | $\approx 80$ | $\approx 30$ |
| Magnetic Pressure $(nPa)$ | $\approx 220$ | $\approx 220$ |
| Plasma Beta $(\beta)$ | $\approx 0.36$ | $\approx 0.13$ |
| Plasma Velocity, $v\ (km/s)$ | $\approx 500$ | $\approx 500$ |
| Alfvén Velocity, $v_A\ (km/s)$ | $\approx 500$ | $\approx 500$ |
| Ion Gyro-radius $(km)$ | $\approx 2$ | $\approx 2$ |
| Ion Inertial Length $(km)$ | $\approx 10$ | $\approx 10$ |
| Plasma Density, $n_{p,i}\ (m^{-3})$ | $\approx 4.5 \times 10^9$ | $\approx 1.25 \times 10^9$ |
| Plasma Temperature, $T_{p,i}\ (K)$ | $\approx 1.5 \times 10^6$ | $\approx 1.3 \times 10^6$ |
| Plasma Frequency, $f_{pe}\ (Hz)$ | $\approx 4 \times 10^5$ | $\approx 3 \times 10^5$ |
| Sound Speed, $Cs\ (km/s)$ | $\approx 130$ | $\approx 130$ |
| Mach Number $(M)$ | $\approx 3$ | $\approx 3.5$ |
| Alfvén Mach Number $(M_A)$ | $\approx 2$ | $\approx 1.1$ |

## 4. Summary

This study presents the observations of Alfvén solitons in the solar corona using data from the Parker Solar Probe (PSP) and the Solar and Heliospheric Observatory (SOHO). The analysis focuses on an interval from 18:09:00 UT to 18:13:00 UT on March 17, 2023, just before the eruption of a Coronal Mass Ejection (CME) detected by SOHO/LASCO C2 coronagraph. The FIELDS instrument onboard PSP provided high-cadence magnetic field data, while plasma parameters have been retrieved from the SWEAP instrument suite. The spacecraft position has been tracked using Helio1-day position data. To identify the solitary structures, Minimum Variance Analysis (MVA) has been applied to the magnetic field data, confirming a train of Alfvén solitons with increasing amplitudes. Additionally, wavelet transform (WT), and power spectral density (PSD) verified their nonlinear nature. The plasma beta parameter remained low during this interval, indicating a magnetically dominated plasma environment, which is favorable for the formation of nonlinear Alfvén wave structures.

A key finding of this investigation is the determination of high Mach numbers associated with these aperiodic solitons. The calculated sonic Mach number exceeds 3, while the Alfvén Mach number is greater than 1.5, indicating that these structures are propagating at *hypersonic* as well as *super-Alfvénic* speeds relative to the background plasma before dissipating their energy to the CME, which is evident from the PSD plots. We emphasize that these magnetic field oscillations, which are found to be a *hypersonic* and *super-Alfvénic* aperiodic train of Alfvén solitons, play a crucial role in transferring energy to the CME, facilitating its expansion and eventual eruption. This solitary train exhibits dissipative characteristics, as evident from the PSD analysis, which demonstrates significant energy transfer to the surrounding plasma. This dissipation mechanism suggests that these structures may contribute to the localized heating of the corona and play a crucial role in the initiation or amplification of CMEs by injecting energy into the surrounding plasma via a process called soliton radiation. Moreover, the PSP's positional data confirmed their presence deep within the corona (0.06 AU from the Sun), the region where Alfvénic turbulence is expected to play a key role in the coronal dynamics. The observed fluctuations in the magnetic field strength, plasma pressure and plasma beta distinctly mark the transition from pre-CME plasma to CME-driven shock, reinforcing the argument that the formation of a train of aperiodic solitons with *hypersonic* and *super-Alfvénic* speeds, is a fundamental mechanism facilitating energy transfer to the plasma.

## 5. Discussion

The detection of Alfvén solitons in the solar corona provides strong observational evidence that nonlinear plasma structures play a significant role in governing the coronal dynamics and CME evolution.

**Key finding include:**

**1. Identifications of Alfvén Solitons:** The MVA method reveal a train of solitons with increasing amplitudes, which are further confirmed by wavelet transform (WT) and power spectral density (PSD) analysis, depicting the nonlinear wave interactions in the coronal plasma.

**2. Plasma conditions:** The presence of low plasma beta ($\beta$) and significant magnetic field fluctuations indicate a highly nonlinear plasma regime, thus providing the apt environment for Alfvén solitons to evolve and propagate.

**3. CME-Soliton Relationship**: The close temporal proximity of the observed solitons to the CME onset demonstrates their significant contribution in facilitating energy transfer mechanisms within the coronal plasma, thereby, initiating the CME as well as contributing to the coronal heating.

**4. Mach Number and Hypersonic/super-Alfvénic Propagation**: The observed train of solitons exhibit a Mach number > 3 and an Alfvén Mach number > 1.5, classifying them as *hypersonic* and *super-Alfvénic* structures (Das, Khusroo, & Bora, 2025). This suggests that they can contribute to high-speed energy transport in the corona, potentially injecting energy into the plasma to develop into a CME, thereby, influencing solar wind acceleration and turbulence dissipation.

Finally, we conclude that these findings have implications for understanding the coronal heating problem, solar wind acceleration, and solar eruptive events like CMEs. However, a comprehensive understanding of the underlying mechanisms governing soliton formation and their role in energy transport requires further theoretical investigation and numerical simulations, which are beyond the scope of this study. Researchers are strongly encouraged to conduct future studies, both theoretical and observational, including advanced simulations and in-situ observations, using the PSP and upcoming heliospheric missions to further explore the impact of solitons in space plasma dynamics.

## 6. Acknowledgement

The authors would like to acknowledge the NASA Parker Solar Probe (PSP) mission and the SWEAP team led by Dr. Justin C. Kasper as well as the FIELDS team led by Prof. Stuart D. Bale for providing the data the used in this study. The SOHO/LASCO data used here are produced by a consortium of the Naval Research Laboratory (USA), Max-Planck-Institut für Aeronomie (Germany), Laboratoire d'Astronomie (France), and the University of Birmingham (UK). SOHO is a project of international cooperation between ESA and NASA. Data analysis was done with the QSAS science analysis system provided by the United Kingdom Cluster Science Centre (Imperial College London and Queen Mary, University of London) supported by the Science and Technology Facilities Council (STFC), along with Python programming using the Jupyter interface. One of the authors, MK, gratefully acknowledges Dr. Steve Schwartz for his invaluable guidance in operating the QSAS with the

PSP data. The author also sincerely appreciates Dr. Tony Allen and Prof. Stuart D. Bale for their valuable assistance in data processing. The author expresses deep gratitude to Dr. Aveek Sarkar, Physical Research Laboratory (PRL), Ahmedabad, India for his invaluable guidance and feedback throughout the research. MK sincerely acknowledges Dr. Madhurjya P. Bora, Department of Physics, Gauhati University, India for his profound insights into soliton dynamics, which have significantly contributed to the advancements of this research. Additionally, the author extends appreciation to Dr. Pallabi Pathak, Department of Physics, University of Science and Technology Meghalaya (USTM), India for her insightful feedback during the application of one of the data analysis techniques.